\documentclass[fleqn,12pt,twoside]{article}
\usepackage[headings]{espcrc1}

\readRCS
$Id: espcrc1.tex,v 1.2 2004/02/24 11:22:11 spepping Exp $
\usepackage{graphicx}

\newcommand{\AmS}{{\protect\the\textfont2
  A\kern-.1667em\lower.5ex\hbox{M}\kern-.125emS}}

\title{Correlations and Fluctuations over a Broad Range in Pseudorapidity 
using the PHOBOS Detector}

\runauthor{G.S.F. Stephans for the PHOBOS collaboration}
\author{G.S.F. Stephans for the PHOBOS collaboration\\
B.Alver$^4$,
B.B.Back$^1$,
M.D.Baker$^2$,
M.Ballintijn$^4$,
D.S.Barton$^2$,
R.R.Betts$^6$,
A.A.Bickley$^7$,
R.Bindel$^7$,
A.Budzanowski$^3$,
W.Busza$^4$,
A.Carroll$^2$,
Z.Chai$^2$,
V.Chetluru$^6$,
M.P.Decowski$^4$,
E.Garc\'{\i}a$^6$,
T.Gburek$^3$,
N.George$^2$,
K.Gulbrandsen$^4$,
S.Gushue$^2$,
C.Halliwell$^6$,
J.Hamblen$^8$,
G.A.Heintzelman$^2$,
C.Henderson$^4$,
I.Harnarine$^6$,
D.J.Hofman$^6$,
R.S.Hollis$^6$,
R.Ho\l y\'{n}ski$^3$,
B.Holzman$^2$,
A.Iordanova$^6$,
E.Johnson$^8$,
J.L.Kane$^4$,
N.Khan$^8$,
W.Kucewicz$^6$,
P.Kulinich$^4$,
C.M.Kuo$^5$,
W.Li$^4$,
W.T.Lin$^5$,
C.Loizides$^4$,
S.Manly$^8$,
A.C.Mignerey$^7$,
R.Nouicer$^{2,6}$,
A.Olszewski$^3$,
R.Pak$^2$,
I.C.Park$^8$,
C.Reed$^4$,
L.P.Remsberg$^2$,
M.Reuter$^6$,
E.Richardson$^7$,
C.Roland$^4$,
G.Roland$^4$,
L.Rosenberg$^4$,
J.Sagerer$^6$,
P.Sarin$^4$,
P.Sawicki$^3$,
I.Sedykh$^2$,
W.Skulski$^8$,
C.E.Smith$^6$,
M.A.Stankiewicz$^2$,
P.Steinberg$^2$,
G.S.F.Stephans$^4$,
A.Sukhanov$^2$,
A.Szostak$^2$,
J.-L.Tang$^5$,
M.B.Tonjes$^7$,
A.Trzupek$^3$,
C.Vale$^4$,
G.J.van~Nieuwenhuizen$^4$,
S.S.Vaurynovich$^4$,
R.Verdier$^4$,
G.I.Veres$^4$,
P.Walters$^8$,
E.Wenger$^4$,
D.Willhelm$^2$,
F.L.H.Wolfs$^8$,
B.Wosiek$^3$,
K.Wo\'{z}niak$^3$,
A.H.Wuosmaa$^1$,
S.Wyngaardt$^2$,
B.Wys\l ouch$^4$\\
\small
$^1$~Argonne National Laboratory, Argonne, IL 60439-4843, USA\\
$^2$~Brookhaven National Laboratory, Upton, NY 11973-5000, USA\\
$^3$~Institute of Nuclear Physics PAN, Krak\'{o}w, Poland\\
$^4$~Massachusetts Institute of Technology, Cambridge, MA 02139-4307, USA\\
$^5$~National Central University, Chung-Li, Taiwan\\
$^6$~University of Illinois at Chicago, Chicago, IL 60607-7059, USA\\
$^7$~University of Maryland, College Park, MD 20742, USA\\
$^8$~University of Rochester, Rochester, NY 14627, USA\\
}
\begin{document}

% typeset front matter
\maketitle

\begin{abstract}
The PHOBOS apparatus includes a charged particle multiplicity detector
covering almost all of 4$\pi$ in solid angle.  The broad coverage in
pseudorapidity, $\eta$, which is unique at RHIC, combined with the ability to collect
a large sample of events with minimal bias, makes possible a study of correlations
and fluctuations over 
%close to the entire 
most of the
pseudorapidity range, even at the highest
beam energy.  Long range correlations, short range correlations at large
$|\eta|$, event-by-event fluctuations in the total number of charged
particles, and event-by-event variation in the full shape of the
pseudorapidity distribution can all be studied.
Preliminary results for the first exploration of these capabilities of the
PHOBOS detector, along with some average properties of the pseudorapidity 
distributions, are presented.
\end{abstract}

\section{Introduction} 
Studies of correlations and fluctuations can reveal a wealth of detailed
information concerning the underlying mechanism of particle production in
collisions of both heavy ions and simpler systems.  
The PHOBOS multiplicity detector consists of single layer Si pad detectors 
supported in an octagonal frame surrounding the interaction region as well as 
in ring-shaped frames arranged along the beam pipe \cite{PhobSi}.  
Additional details of the detector layout as well as information on the
triggering and centrality selections for the various systems studied can
be found in \cite{PhobWP}.  The uniquely broad
pseudorapidity ($\eta$) coverage of the PHOBOS multiplicity detector 
($|\eta|\leq5.4$) 
enables a
wide variety of studies of both correlations and fluctuations.  This paper
presents the results of a few preliminary studies exploiting these unique
capabilities.

\section{Two particle correlations}
\begin{figure} [ht]
\vspace{-0.5cm}
\begin{minipage}[t]{7.0cm}
\vspace{-3.5cm}
In a small system like d+Au, this
uncorrected correlation function is dominated by detector effects such as
$\delta$-electrons which cause a peak at small $\Delta\eta$ and $\Delta\phi$
and momentum conservation which causes a ridge at small $\Delta\eta$ and
$\Delta\phi\approx180\deg$.  
Work is ongoing to correct for these
uninteresting effects and study the remaining correlations in the data.
\end{minipage}
\hspace{\fill}
\begin{minipage}[t]{8.0cm}
\begin{center}
\includegraphics[width=8cm]{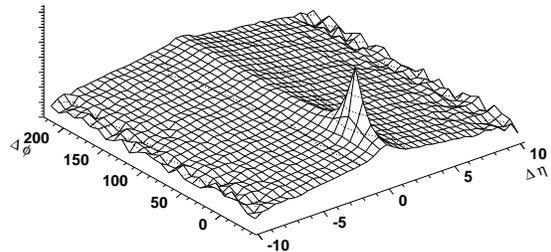}
\end{center}
\vspace{-1.5cm}
\caption{Uncorrected two particle correlation function measured in 
collisions
of d+Au at $\sqrt{s_{_{NN}}}=200$~GeV. See text for discussion.}
\label{dAuCorrelRaw}
\end{minipage}
\vspace{-0.5cm}
\end{figure}

The utility of broad $\eta$ coverage is illustrated in Fig.~\ref{dAuCorrelRaw}
which shows the raw two-particle correlation function measured for
%minimum-bias 
collisions of d+Au at $\sqrt{s_{_{NN}}}=200$~GeV.  It is clear
that these data can be used to study both correlations of particles widely
separated in $\eta$ as well as particles closely separated in $\eta$ but far
from mid-rapidity.  
The histogram shown in Fig.~\ref{dAuCorrelRaw} was
obtained by counting all pairs of particles within an event in bins of $\Delta\eta$
and $\Delta\phi$ and dividing by the same binning of randomly selected pairs of
particles from different events.  

\section{Properties of the pseudorapidity distributions}
\begin{figure}[h!]
\begin{minipage}[t]{7.75cm}
\begin{center}
\includegraphics[width=7.75cm]{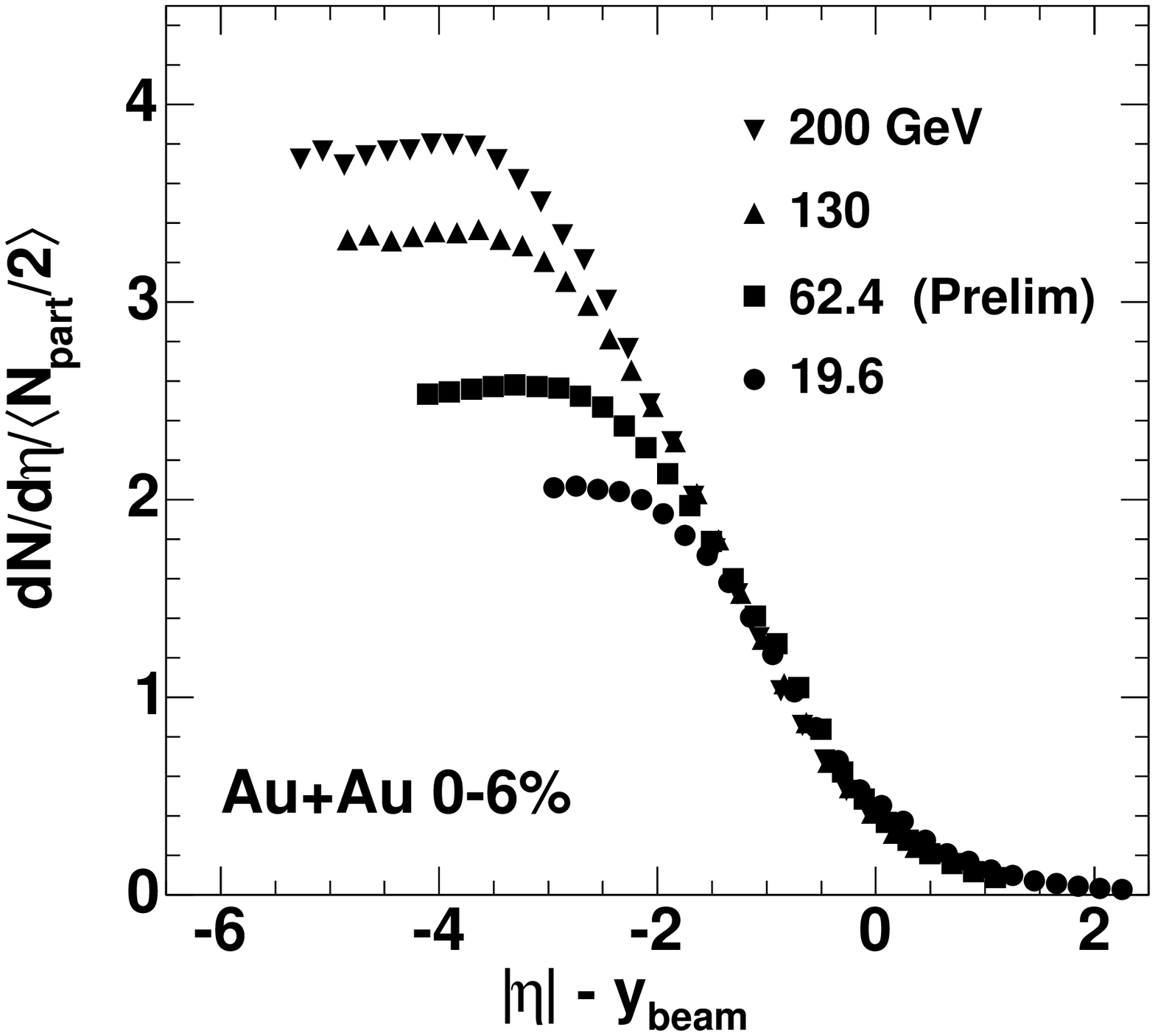}
\end{center}
\vspace{-1.5cm}
\caption{Pseudorapidity density distributions of charged particles emitted in
central collisions of Au+Au at four energies plotted in the approximate rest
frame of one of the colliding nuclei \cite{ELS}.  See text for
discussion.}
\label{ExtLongScaleAuAu}
\end{minipage}
\hspace{\fill}
\begin{minipage}[t]{7.75cm}
\begin{center}
\includegraphics[width=7.75cm]{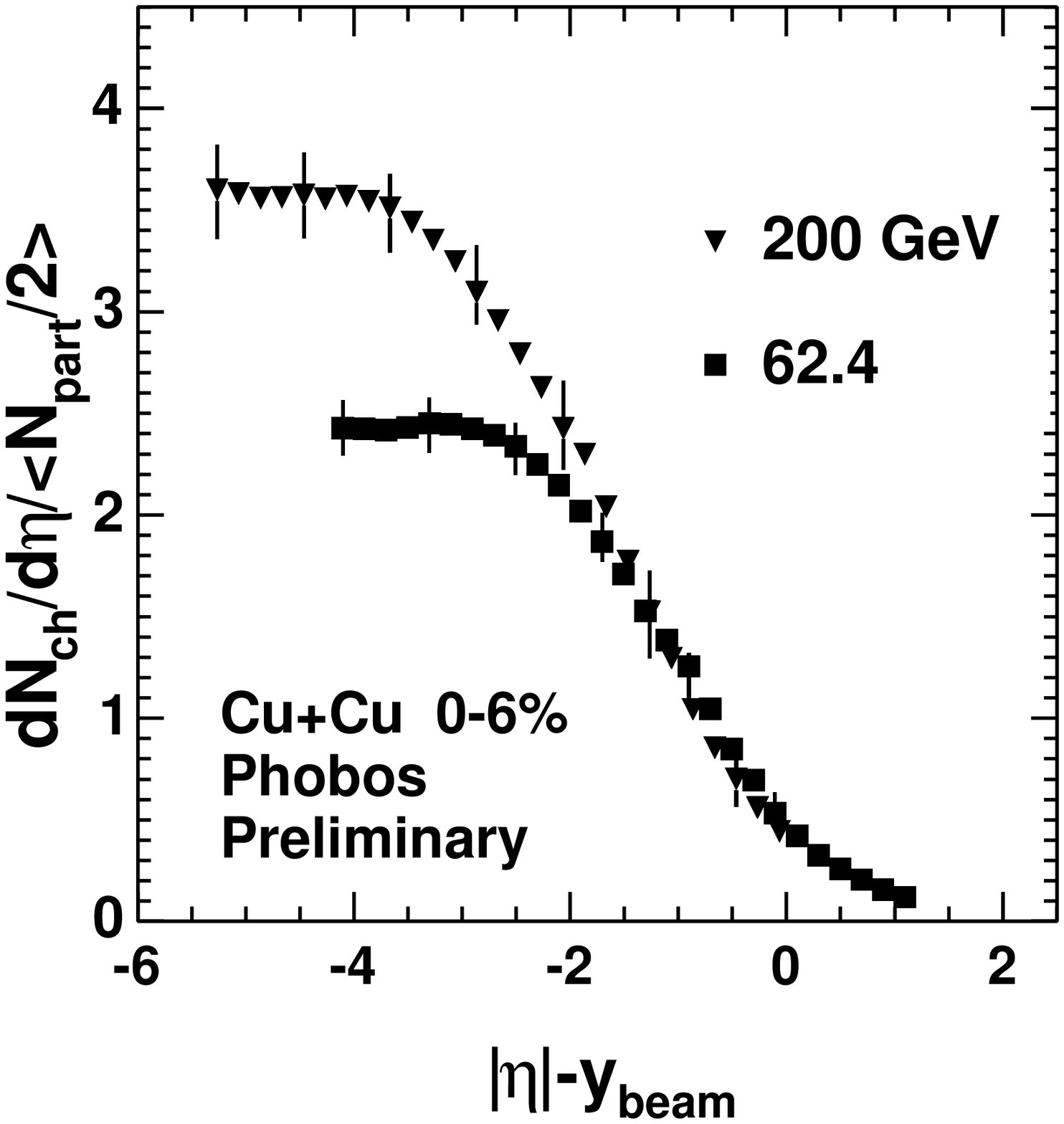}
\end{center}
\vspace{-1.5cm}
\caption{Pseudorapidity density distributions of charged particles emitted in
central collisions of Cu+Cu at two energies plotted in the approximate rest
frame of one of the colliding nuclei.  See text for
discussion.}
\label{ExtLongScaleCuCu}
\end{minipage}
%\end{figure}
%\begin{figure}[ht]
\begin{center}
\includegraphics[width=8.5cm]{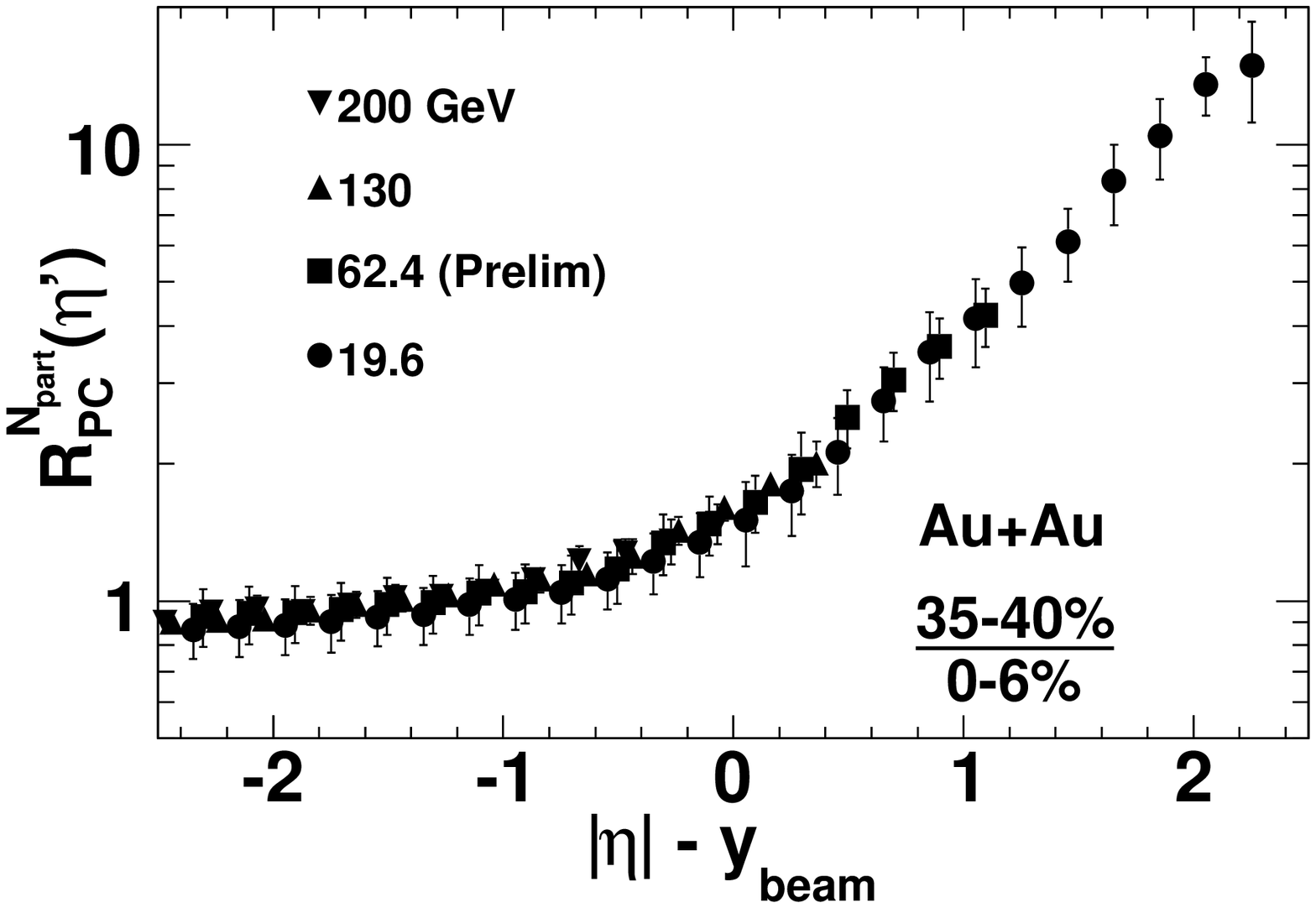}
\end{center}
\vspace{-1.5cm}
\caption{The ratio of dN/d$\eta$ distributions of charged particles for
semi-peripheral over central collisions for Au+Au at four energies, normalized
by the number of participating nucleons, plotted in the approximate rest
frame of one of the colliding nuclei \cite{ELS}.  See text for
discussion.}
\label{EndCentFactAuAu}
\vspace{-1.4cm}
\end{figure}

Before discussing the fluctuations in the distributions of charged particles,
it is useful to review the properties of the distributions averaged over all
events. Fig.~\ref{ExtLongScaleAuAu} summarizes the dominant feature of such
distributions, namely extended longitudinal scaling.  Distributions of
dN/d$\eta$ for charged particles emitted in central collisions of Au+Au at
four energies are shown \cite{ELS}.  The data at positive and negative
$\eta$ have been averaged and then plotted versus $|\eta|-y_{beam}$, thereby
approximating 
%viewing them in 
the rest frame of one of the colliding nuclei.
The distribution of charged particles, when viewed in this rest frame, is
clearly independent of energy over a large range of longitudinal phase space.
Preliminary results from the recent Cu+Cu run show identical behavior (see
Fig.~\ref{ExtLongScaleCuCu}).

The distributions for non-central collisions at a given energy differ from
those of central collisions, both in shape and height but all show the same
beam energy scaling. The degree to which the energy and centrality
dependencies factorize is shown in Fig.~\ref{EndCentFactAuAu}.  The dN/d$\eta$
distributions for collisions in the centrality bin 35$-$40\% are normalized by
the number of participating nucleons and divided by the similarly normalized
distribution for central collisions.  Again, the data are plotted in the
approximate rest frame of one of the colliding nuclei.  The variation in
shape (represented by the large deviations from a constant value) are seen to
be remarkably independent of energy over the entire range measured.

\section{Fluctuations in the dN/d$\eta$ distributions}
\begin{figure}[ht]
\begin{minipage}[t]{7.75cm}
\begin{center}
\includegraphics[width=7.75cm]{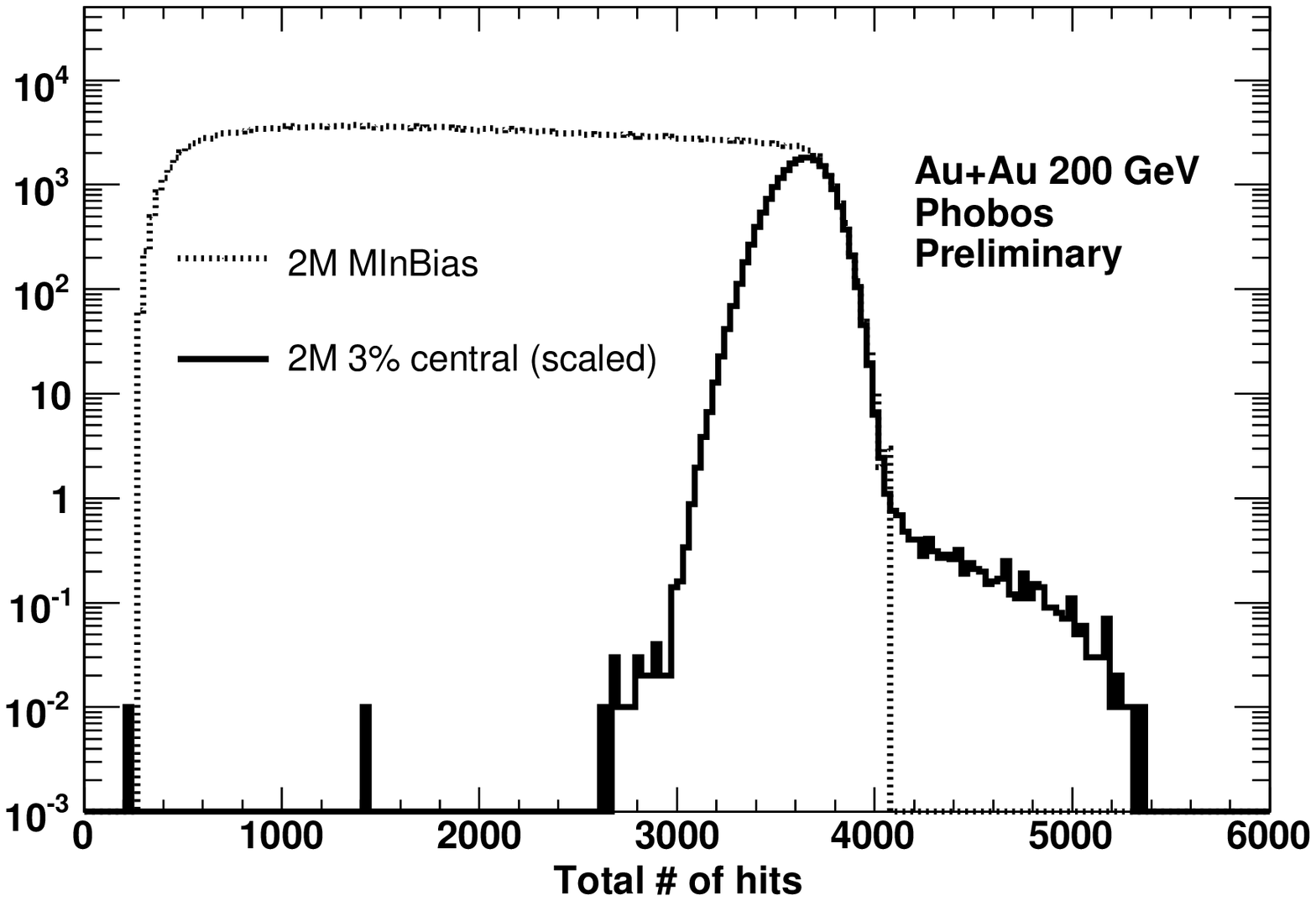}
\end{center}
\vspace{-1.5cm}
\caption{The distribution of the total number of detected charged particles 
emitted in
%minimum bias 
events without centrality selection and in the 3\% most central Au+Au events.}
\label{BigMult}
\end{minipage}
\hspace{\fill}
\begin{minipage}[t]{7.75cm}
\begin{center}
\includegraphics[width=7.75cm]{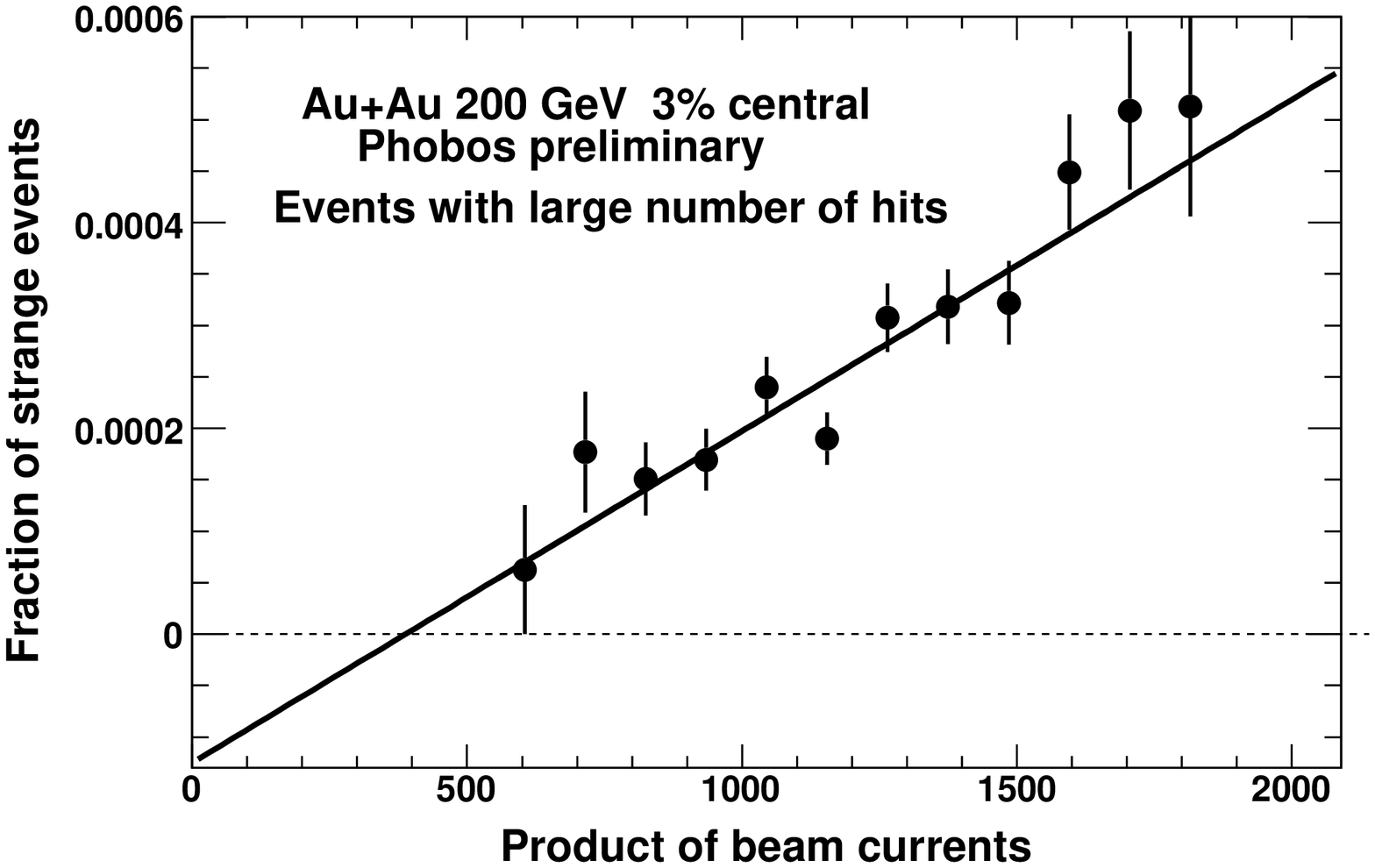}
\end{center}
\vspace{-1.5cm}
\caption{The number of events with more than 4400 detected charged particles
versus the product of the currents in the two RHIC beams.}
\label{BigPileup}
\end{minipage}
\vspace{-0.5cm}
\end{figure}

Having established the average properties of the dN/d$\eta$ distributions, two
preliminary studies of fluctuations in the distributions were performed.  Data
for the most central 3\% Au+Au collisions from the high statistics run at
$\sqrt{s_{_{NN}}}$=200~GeV were used, a total of about 2 million events.  The
first investigation looked for events with unusually high total number of
charged particles.  The results are shown in Fig.~\ref{BigMult}.  A clear, 
but small, tail
extending to high multiplicity is evident.  
A preliminary conclusion can be
drawn that events with larger than expected total multiplicity are rarer than
a few times $10^{-4}$. One obvious source of such events
is pile-up in which two collisions occur within the read-out time of the
experiment.  In order to study this possibility, the rate of events with a
number of hits above 4400 (about 570 in total) was plotted versus a crude
measure of the event rate, with the results shown in Fig.~\ref{BigPileup}.  A
clear correlation is evident which extrapolates to zero events at zero
luminosity to within less than 2$\sigma$.
% and possibly one or two orders of magnitude smaller.

An additional study was performed to look for events with unusual shapes of
the dN/d$\eta$ distribution.  Events from the same central Au+Au sample were
used.  Within bins in vertex location (to eliminate trivial acceptance
effects), the average dN/d$\eta$ shape and variance was determined.  Each
event was normalized to the same total number of hits and then a $\chi^2$ was
calculated comparing to the average event shape.  A clear excess of events
with large $\chi^2$ was observed, about 200 in total.  In this case, a 
slightly smaller preliminary
limit of about $1\times10^{-4}$  is
found.  However, as with the
high multiplicity events, the number of these unusual events was correlated 
with the
event rate with an extrapolation close to zero, suggesting the possibility of 
smaller values.

\section {Summary}
The unique properties of the PHOBOS charged particle multiplicity detector are
being exploited in the study of correlations and fluctuations in the
production of particles in collisions at RHIC.  Average pseudorapidity density
distributions in Au+Au and preliminary Cu+Cu data show extended longitudinal
scaling and a 
%dramatically perfect 
striking 
factorization of the dependencies on energy and
centrality.  Preliminary searches for events with unusually high total number
of charged particles and with unusual shapes in dN/d$\eta$ have yielded small
numbers (frequencies on the order of $10^{-4}$) with strong indications that a
large fraction, possibly close to all of the events, are due to simple
pileup.  Work is ongoing to determine more precise limits for these two
classes of unusual events.

\end{document}